\documentclass[aps, prb, twocolumn]{revtex4-2}

\usepackage{enumerate}
\usepackage{comment}
\usepackage{amsmath}
\usepackage{amssymb}
\usepackage{color}
\usepackage{xcolor}
\usepackage{tikz}
\usepackage{amsthm}
\usepackage{graphicx}
\usepackage{pict2e}
\usepackage{bbold}
\usepackage{empheq}
\usepackage{subfigure}
\usepackage{xfrac}
\usepackage[colorlinks,citecolor=blue,linkcolor=blue,urlcolor=blue,bookmarks=false,hypertexnames=true]{hyperref}
\usepackage{enumitem}
\usepackage{notes2bib}

\newcommand{\RD}{\rangle}
\newcommand{\Hh}{\hat{\text{H}}}
\newcommand{\HLR}{\hat{ {\Lambda}}}
\newcommand{\VLR}{{ {\Lambda}}}
\newcommand{\HSR}{\hat{ { \Upsilon}}}
\newcommand{\VSR}{{ { \Upsilon}}}
\newcommand{\Hhcont}{\Hh_{\rm cont}}
\newcommand{\Hhhub}{\Hh}

\newcommand{\SPorbital}{\mathfrak{i}}

\newcommand{\SPspin}{\mathfrak{s}}
\newcommand{\SPvalley}{\mathfrak{t}}

\newcommand{\Couplg}{\mathcal{G}}

\newcommand{\ccdag}{\text{c}^{\dagger}}
\newcommand{\cc}{\text{c}}

\newcommand{\eqn}{Eq.~}
\newcommand{\eqns}{Eqs.~}
\newcommand{\fig}{Fig.~}
\newcommand{\figs}{Figs.~}

\newcommand{\reffs}{Refs.~}

\newcommand{\SPState}{\Psi }

\newcommand{\gVSP}{g_{v}^{\SPorbital}}

\newcommand{\TP}{two-particle}
\newcommand{\SP}{single-particle}
\newcommand{\SO}{spin-orbit}

\newcommand{\BLG}{BLG}

\newcommand{\GS}{ground state}

\newcommand{\QD}{QD}
\newcommand{\DD}{double-dot}
\newcommand{\DQD}{double-QD}

\definecolor{gold}{HTML}{D6A800}
\definecolor{taupe}{HTML}{BC80BD}

\begin{document}

\title{Extended Hubbard model describing small multi-dot arrays in  bilayer graphene}

\author{Angelika Knothe$^{1}$ and Guido Burkard$^{2}$}
\affiliation{$^1$Institut f\"ur Theoretische Physik, Universit\"at Regensburg, D-93040 Regensburg, Germany}
\affiliation{$^2$Department of Physics, University of Konstanz, D-78457 Konstanz, Germany}
\date{\today}


\begin{abstract}
{We set up and parametrize a Hubbard model for interacting quantum dots in bilayer graphene and study double dots as the smallest multi-dot system. We demonstrate the tunability of the spin and valley multiplets, Hubbard parameters, and effective exchange interaction by electrostatic gate potentials and the magnetic field. Considering both the long- and short-range Coulomb interactions, we map out the various spin and valley multiplets and calculate their energy gaps for different dot sizes and inter-dot separations. For half-filling and large valley splittings, we derive and parametrize an effective Heisenberg model for the quantum dot spins.
}
\end{abstract}
 
\maketitle

\section{Introduction}

Gate-defined nanostructures in gapped bilayer graphene (BLG) have emerged as promising platforms for quantum confinement and quantum technologies. Recent advancements in experimental techniques have enabled the successful manipulation and characterization of quantum dots (QDs) in \BLG{} \cite{eichSpinValleyStates2018, banszerusGateDefinedElectronHole2018, garreisCountingStatisticsSingle2023, garreisShellFillingTrigonal2021, tongTunableValleySplitting2021, kurzmannKondoEffectSpin2021, gachterSingleShotSpinReadout2022, banszerusDispersiveSensingCharge2020, banszerusElectronHoleCrossover2020, arXiv:2305.09284, arXiv:2303.10201, mollerProbingTwoElectronMultiplets2021, banszerusPulsedgateSpectroscopySingleelectron2020, banszerusSpinRelaxationSingleelectron2022, banszerusSpinvalleyCouplingSingleelectron2021, renRealizingValleyPolarizedEnergy2022, banszerusTunableInterdotCoupling2021, Jing_2023, duprez2023spectroscopy}. These achievements range from identifying the single-particle  \cite{eichSpinValleyStates2018,banszerusGateDefinedElectronHole2018, garreisShellFillingTrigonal2021, duprez2023spectroscopy} and two-particle states \cite{mollerProbingTwoElectronMultiplets2021, arXiv:2305.09284} in a \BLG{} \QD{} to the measurement of spin 
 \cite{banszerusSpinRelaxationSingleelectron2022, gachterSingleShotSpinReadout2022, banszerusPulsedgateSpectroscopySingleelectron2020} and valley \cite{arXiv:2304.00980} lifetimes. To scale up BLG QD systems for quantum information setups and to fully understand the  material-dependent features observed in the \QD{}s, it is necessary to develop a theoretical framework encompassing multi-dot systems.

\begin{figure}[t!]
    \centering
    \includegraphics[width=0.92\linewidth]{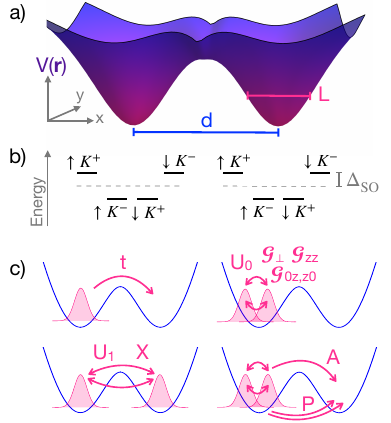}
    \caption{a) Two-dimensional confinement potential $V(\mathbf{r})$ for a \DQD{} with dot diameter $L$ and inter-dot separation $d$. b) The $B=0$ \SP{} levels  comprise four spin ($\uparrow, \downarrow$) and valley ($K^{\pm}$) states per dot, separated by the \SO{} coupling gap $\Delta_{\rm SO}$. A finite perpendicular magnetic field will split these levels further proportional to their spin- and valley g-factors. c) Single- and \TP{} processes:  tunnelling ($t$), on-site interactions ($\propto U_0, \Couplg_{\perp, zz, 0z, z0} $), nearest neighbour direct ($U_1$) and exchange (X) interaction, density-assisted hopping ($A$) and pair-hopping ($P$).}
    \label{fig:1}
\end{figure}

In this work, we address this  need by setting up a generalized Fermi-Hubbard Hamiltonian for multiple interacting \BLG{} \QD{}s. Building upon  successful microscopic models developed for individual \QD{}s \cite{knotheQuartetStatesTwoelectron2020, garreisShellFillingTrigonal2021, mayerTuningConfinedStates2023, tongTunableValleySplitting2021, mollerProbingTwoElectronMultiplets2021, knotheTunnelingTheoryBilayer2022, arXiv:2305.09284, PhysRevB.78.195427, recherBoundStatesMagnetic2009}, we parametrize the Hamiltonian to capture the essential system characteristics. We then study the low-energy states and Hubbard parameters of \DQD{}s, as illustrated in \fig{}\ref{fig:1}.

The theoretical description of multi-dot systems is paramount for comprehending the recent and ongoing experiments in the field \cite{eichCoupledQuantumDots2018, arXiv:2305.03479, arXiv:2303.10119, arXiv:2304.00980, arXiv:2211.04882, banszerusSingleElectronDoubleQuantum2020, tongPauliBlockadeTunable2022}. Furthermore, the investigation of coupled \QD{}s holds promise for spin and valley qubit formation and control \cite{rohlingUniversalQuantumComputing2012, schaibleyValleytronics2DMaterials2016}, which are essential elements in quantum computing \cite{Burkard2023,lossQuantumComputationQuantum1998a,burkardCoupledQuantumDots1999,fernandez-fernandezQuantumControlHole2022, brooksHybridExchangeMeasurementBased2021, heinzCrosstalkAnalysisSimultaneously2022, arXiv:2303.17211}. Beyond these immediate applications, our research also explores an extended Hubbard model incorporating long-range interactions and diverse spin and valley states that can be tuned by adjusting inter-dot separation and magnetic fields.

The development of such an extended model paves the way for describing larger \BLG{} \QD{} lattices and the use of \BLG{} \QD{}s for quantum simulations of exotic Fermi-Hubbard Hamiltonians. Extended Hubbard models have been shown to potentially host diverse correlation effects such as magnetism \cite{readFeaturesPhaseDiagram1989, tamuraFerromagnetismDDimensionalSU2021, samajdar2023polaronic, willsherMagneticExcitationsPhase2023, yuMagneticPhasesAnisotropic2023, dehollainNagaokaFerromagnetismObserved2020, arXiv:2305.05683, arXiv:2304.04563, arXiv:2303.08324, buterakos2023magnetic, kapcia2023magnetic, feng2023metalinsulator, Gall_2021}, superconducting pairing \cite{affleckLargeLimitHeisenbergHubbard1988, gilmutdinovInterplayMagnetismSuperconductivity2022, arXiv:2206.01119, arXiv:2211.06498, kundu2023cdmfthfd}, Kondo physics \cite{arXiv:2305.06734}, localisation effects \cite{freyHilbertSpaceFragmentation2022, arXiv:2301.08246},  translational and rotational symmetry breaking phases \cite{huangStripeOrderPerspective2018, philoxeneSpinChargeModulations2022, arXiv:2304.04563, arXiv:2304.08683, gebhard2023generic, botzung2023exact}, and various topological states \cite{arXiv:2301.03312, arXiv:2301.03312, dvirRealizationMinimalKitaev2023, arXiv:2306.07696, MANSOURY2023129115}.

Successful realizations of multi-dot systems using semiconductor \QD{}s demonstrate their potential, e.g., for quantum simulation \cite{salfiQuantumSimulationHubbard2016, duskoAdequacySiChains2018, vandiepenQuantumSimulationAntiferromagnetic2021},  quantum state transfer and routing \cite{kandelAdiabaticQuantumState2021, arXiv:2209.00920, millsShuttlingSingleCharge2019, yonedaCoherentSpinQubit2021, utsugi2023single}, and inducing extended interactions \cite{qiaoLongDistanceSuperexchangeSemiconductor2021} in 1D \cite{volkLoadingQuantumdotBased2019, gaudreauStabilityDiagramFewElectron2006, reedReducedSensitivityCharge2016, meyerElectricalControlUniformity2023, arXiv:2208.11784, cifuentes2023impact} and 2D \cite{mortemousqueCoherentControlIndividual2021, vandiepenElectronCascadeDistant2021, ansaloniSingleelectronOperationsFoundryfabricated2020, leExtendedHubbardModel2017, lawrieQuantumDotArrays2020, wangExperimentalRealizationExtended2022, mortemousqueEnhancedSpinCoherence2021, throckmortonCrosstalkChargenoiseinducedMultiqubit2022, borsoiSharedControl162023} \QD{} arrays. Here, we open up alternative avenues for quantum simulation, quantum computation, and exploring emergent phenomena using confined systems in materials with complex electronic structures and enriched degrees of freedom, such as gapped \BLG{}.

Analyzing the Hubbard model for \BLG{} \QD{} lattices and a double-dot as the minimal realisation, we establish:
\begin{itemize}[align=right,itemindent=2em,labelsep=2pt,labelwidth=1em,leftmargin=0pt,nosep]
    \item the low-energy states, comprising different spin and valley multiplets, whose multiplicities and splittings depend on the dot parameters,
\item a microscopic understanding how different phases are driven by the competition of long-range extended Hubbard parameters and short-range interactions,
    \item  an effective spin-spin Heisenberg model describing the low-energy double dot states in a magnetic field, whith a tunable effective exchange constant.
\end{itemize} 
These findings will help understanding the on-going experiments based on few-electron tunnel spectroscopy, identifying two-level regimes suitable for forming qubits, and scaling \BLG{} \QD{}s to larger lattices for quantum information and quantum simulation. 

We obtain the results above by setting up  a generalised Hubbard model for  multiple \BLG{} \QD{}s \cite{affleckLargeLimitHeisenbergHubbard1988, campbellBondchargeCoulombRepulsion1988a, campbellModelingElectronelectronInteractions1990, kivelsonMissingBondchargeRepulsion1987, yangGenericHubbardModel2011},
\begin{equation}
\Hhhub=\sum_{j,k=1}^{N}t_{jk}\ccdag_{j}\cc_{k}+\frac{1}{2}\sum_{h,j,k,m}U_{hjkm}\,\ccdag_{h}\ccdag_{j}\cc_{k}\cc_{m},
\label{eqn:HHub}
\end{equation}
where $N$ denotes the total number of \SP{} states. We  parametrise this Hubbard model by evaluating the Hubbard parameters (cf.~\fig{}\ref{fig:1}),
\begin{align}
\nonumber t_{jk}& =\langle j  |\hat H_0| k \rangle,\\
U_{hjkm}&=\langle h j |\hat H_{C}| k m\rangle,
\label{eqn:Hubgen}
\end{align}
in terms of the generalised, orthonormalised \SP{} states $|j\rangle$ of the $j$-th lattice site. These \SP{} states we obtain from a microscopic model for an interacting \BLG{} \DQD{}. In our microscopic model, $\Hhcont=\Hh_{0}+\Hh_{C}$,  the \SP{} term $\Hh_{0}$ includes a smoothly varying potential landscape confining electrons at multiple sites (\fig{}\ref{fig:1}  for a \DQD{} with eight \SP{} spin and valley states, hence $N=8$). The term $\Hh_{C}$ captures material-specific electron-electron Coulomb interactions, including long-range isotropic contributions and short-range interactions sensitive to the \BLG{} lattice and valley structure  (We define all terms of the Hamiltonian and the  \SP{} states in detail below). Using exact diagonalisation of the Hubbard Hamiltonian \eqn\eqref{eqn:HHub}, we study the \TP{} spectra and states of small multi-dot systems.

\section{Microscopic Model of the double dot and parametrization of the Hubbard Hamiltonian}
 Smooth confinement of charge carriers in gapped \BLG{} has been achieved experimentally using a combination of multiple gates (typically, two split gates to confine a channel and finger gates crossing the channel on top to confine a dot)\cite{
Overweg2018, leeTunableValleySplitting2020, overwegTopologicallyNontrivialValley2018,  eichCoupledQuantumDots2018, arXiv:2305.03479, arXiv:2303.10119, arXiv:2304.00980, arXiv:2211.04882, banszerusSingleElectronDoubleQuantum2020, tongPauliBlockadeTunable2022}. The gates locally tune both the gap, and the charge carrier density and hence confine the charge carriers electrostatically.  For a \BLG{} \DQD{} as the smallest possible multi-dot system we choose a smooth confinement potential, $V(\mathbf{r})$ and a spatially modulated gap, $\Delta(\mathbf{r})$, of the  form depicted in \fig{}\ref{fig:1} a),
\begin{align}
\nonumber& V(\mathbf{r})=V_{0}\,V_{\rm conf}(\mathbf{r}), \hskip7pt \Delta(\mathbf{r})=\Delta_{0}-\Delta_{\rm mod} \, V_{\rm conf}(\mathbf{r}),\\
\nonumber &\text{where } \\
\nonumber &V_{\rm conf}(\mathbf{r})=\frac{1}{\cosh{\sqrt{({x}-{x}_{i})^{2}+y^{2}}/L}}-V_{\rm neck}({x}_{i}),\\
& V_{\rm neck}=\frac{1}{L^2}\frac{Z}{2 {x}^2_{i}}(x-{x}_{i})^{4}\Theta(|{x}_{i}|-|{x}|),
\label{eqn:Confinement}
\end{align}
with $i=l$ for $x<0$ (left) and  $i=r$ for $x>0$ (right) and $\Theta$ is the Heaviside step function. The parameters $V_0$  and $\Delta_0$ in \eqn{}\eqref{eqn:Confinement} describe the depth of the confinement and the \BLG{} band gap in the absence of any spatial modulation (away from the dots), while the term proportionally to $\Delta_{\rm mod} $ captures the modulation of the gap towards the centre of the dot (for the magnitude of the modulation $\Delta_{\rm mod} \approx0.3\Delta_0$ has  proven realistic in recent theoretical and experimental works \cite{tongTunableValleySplitting2021, knotheQuartetStatesTwoelectron2020, garreisShellFillingTrigonal2021, arXiv:2305.09284}). Moreover, $L$, defines the diameter of the individual \QD{}s, while $d=x_r-x_l   $ is the distance between the dots (see \fig{}\ref{fig:1}). Further, $Z$ is a fit parameter which we choose such that the neck of the potential \cite{yannouleasBarriersDeformationFission1995, yannouleasCouplingDissociationArtificial2001, yannouleasSpontaneousSymmetryBreaking1999, yannouleasStronglyCorrelatedWavefunctions2002, moselFragmentShellInfluencesNuclear1971, mustafaAsymmetryNuclearFission1973a} is smooth at $x=0$. In \BLG{}, skew hopping between the two graphene layers breaks rotational symmetry, distinguishing between the two crystallographic directions. Below, we consider  dots oriented along  the x-direction (corresponding to aligning them along the zigzag direction of the \BLG{} lattice). Orientation of the dots along the y-direction (aligning along the armchair direction) leads to qualitatively similar results which we provide in Appendix~\ref{app:ExtDat}. The potential and gap of \eqn{}\eqref{eqn:Confinement} enter into  the \SP{} four-band BLG Hamiltonian \cite{mccannLowEnergyElectronic2007, mccannElectronicPropertiesBilayer2013},
\begin{widetext}
\begin{align}
 H^{}_{0}(\SPspin,\SPvalley)&\!= \!\!
\setlength{\arraycolsep}{-11pt} \begin{pmatrix} 
  V  -\frac{1}{2}\SPvalley\Delta+ \SPspin\SPvalley \Delta_{\rm SO} + \SPspin g_S \mu_B B & \SPvalley v_3\pi & 0 &\SPvalley v \pi^{\dagger}\\
\SPvalley v_3 \pi^{\dagger}&  V +\frac{1}{2}\SPvalley\Delta + \SPspin\SPvalley \Delta_{\rm SO} + \SPspin g_S \mu_B B   &\SPvalley v\pi &0\\
 0 & \SPvalley v\pi^{\dagger} &   V +\frac{1}{2}\SPvalley\Delta + \SPspin\SPvalley \Delta_{\rm SO} + \SPspin g_S \mu_B B   &   \gamma_1\\
\SPvalley v\pi & 0 &   \gamma_1 &  V -\frac{1}{2}\SPvalley\Delta + \SPspin\SPvalley \Delta_{\rm SO} + \SPspin g_S \mu_B B 
\end{pmatrix},
\label{eqn:H}
\end{align}
\end{widetext}
where $\pi=p_x+ip_y,\,  \pi^{\dagger}=p_x-ip_y,\,  \mathbf{p}=-i\hbar\nabla-\frac{e}{c}\boldsymbol{\mathcal{A}},$ with elementary charge, $e > 0$,  speed of light, c, and vector potential, $\boldsymbol{\mathcal{A}} = \frac{B}{2} (-y,x,0) $  in a symmetric gauge for an out-of plane magnetic field, B. The Zeeman coupling of the two spin states, $\uparrow, \downarrow$ ($\SPspin=\pm1$) is proportional to the spin g-factor, $g_s$, and the Bohr magneton, $\mu_B$. Further, $\Delta_{\rm SO}$ is a Kane-Mele type \SO{}-coupling gap enhanced by zero-point vibrations \cite{PhysRevB.86.245411, banszerusObservationSpinOrbitGap2020, banszerusSpinvalleyCouplingSingleelectron2021, kurzmannKondoEffectSpin2021}, $ v=1.02*10^6 $ m/s, $  v_3\approx0.12 v$, and $ \gamma_1=0.38 $ eV.
The above Hamiltonian is written in the basis $\Phi_{K^+}=(\phi_{A},\phi_{B^{\prime}},\phi_{A^{\prime}},\phi_{B})$ or $\Phi_{K^-}=(\phi_{B^{\prime}},\phi_{A},\phi_{B},\phi_{A^{\prime}})$ of states on the four \BLG{} sub-lattices  in the two valleys, $K^{\pm}$ (indexed by $\SPvalley=\pm1$). Confinement models of similar shape have been used previously successfully to describe individual, single \QD{}s in \BLG{} \cite{knotheQuartetStatesTwoelectron2020, garreisShellFillingTrigonal2021, mollerProbingTwoElectronMultiplets2021, knotheTunnelingTheoryBilayer2022, arXiv:2305.09284, mayerTuningConfinedStates2023}.

For the electron-electron interactions in \eqns{}\eqref{eqn:HHub} and \eqref{eqn:Hubgen},  $\Hh_{C}=\HLR+\HSR$, we take into account the screened long-range Coulomb interaction, $\HLR$, and short-range interactions, $\HSR$, where the latter break sublattice and valley symmetry on the lattice scale. The  short-range interactions stem from symmetry breaking fluctuations and favour states with spontaneously broken symmetries \cite{cheianovGappedBilayerGraphene2012, kharitonovCantedAntiferromagneticPhase2012, kharitonovPhaseDiagramNu2012, lemonikCompetingNematicAntiferromagnetic2012}. The respective corresponding matrix elements, \eqn{}\eqref{eqn:Hubgen}, read,
\begin{widetext}
\begin{align}
\nonumber    & \langle h\!:\!\SPorbital_1\SPspin_{1}\SPvalley_{1},{j}\!:\!\SPorbital_2\SPspin_{2}\SPvalley_{2}|\HLR|{m}\!:\!\SPorbital_4\SPspin_{4}\SPvalley_{4},{k}\!:\!\SPorbital_3\SPspin_{3}\SPvalley_{3}\rangle  
 = \delta_{\SPspin_{1}\SPspin_{4}} \delta_{\SPspin_{2}\SPspin_{3}} \delta_{\SPvalley_{1}\SPvalley_{4}}  \delta_{\SPvalley_{2}\SPvalley_{3}} \! \iint \!\!d\mathbf{r}d\mathbf{r}^{\prime}[\SPState^{{*}}_{h:\SPorbital_1}(\mathbf{r})\SPState^{*}_{j:\SPorbital_2}(\mathbf{r}^{\prime})] \, U_{\rm scr}(\mathbf{r}-\mathbf{r}^{\prime})\,[\SPState_{k:\SPorbital_3}(\mathbf{r}^{\prime})\SPState_{m:\SPorbital_4}(\mathbf{r})],\\
\label{eqn:MLR} &U_{\rm scr}(\mathbf{q})=\frac{e^{2}}{4\pi\epsilon\epsilon_{0}}\frac{2\pi}{q(1+qR_{\star})},\hskip10pt U_{\rm scr}(\mathbf{r})=\int\frac{d^{2}q}{(2\pi)^{2}}e^{i\mathbf{q}.\mathbf{r}}U_{\rm scr}(\mathbf{q}),\\
\nonumber    &\langle h\!:\!\SPorbital_1\SPspin_{1}\SPvalley_{1},{j}\!:\!\SPorbital_2\SPspin_{2}\SPvalley_{2}|\HSR|{m}\!:\!\SPorbital_3\SPspin_{4}\SPvalley_{4},{k}\!:\!\SPorbital_4\SPspin_{3}\SPvalley_{3}\rangle 
 = \delta_{\SPspin_{1}\SPspin_{4}} \delta_{\SPspin_{2}\SPspin_{3}} \delta_{hjmk}\mathfrak{J}_{hjmk}
 \begin{cases} 
\Couplg_{zz}+\Couplg_{z0}+\Couplg_{0z},  & \text{if } \SPvalley_{1}=\SPvalley_{2}= \SPvalley_{4}= \SPvalley_{3} ,\\
\Couplg_{zz}-\Couplg_{z0}-\Couplg_{0z}, & \text{if }  \SPvalley_{1}=\SPvalley_{4}=\SPvalley,  \SPvalley_{3}= \SPvalley_{2} =\SPvalley^{\prime} ,  \SPvalley\neq \SPvalley^{\prime}\\
4\Couplg_{\perp}, & \text{if } \SPvalley_{1}=\SPvalley_{3}=\SPvalley,  \SPvalley_{2}= \SPvalley_{4} =\SPvalley^{\prime} , \SPvalley\neq \SPvalley^{\prime},
\end{cases}\\
&\mathfrak{J}_{{hjkl}}= \int d\mathbf{r} \;\SPState^{{*}}_{ h:\SPorbital_1}(\mathbf{r}) \SPState^{{*}}_{ j:\SPorbital_2}(\mathbf{r})\SPState_{k:\SPorbital_3}(\mathbf{r})\SPState^{{}}_{m:\SPorbital_4}(\mathbf{r}).
\label{eqn:MSR}
\end{align}
\end{widetext}
Here, $|j\!\!:\!\SPorbital \SPspin \SPvalley \rangle  $ denotes the state with orbital quantum number $\SPorbital$, spin $\SPspin $, and valley index $\SPvalley $  of the $j$-th \QD{}.

\begin{figure*}[ht]
    \centering
    \includegraphics[width=1\linewidth]{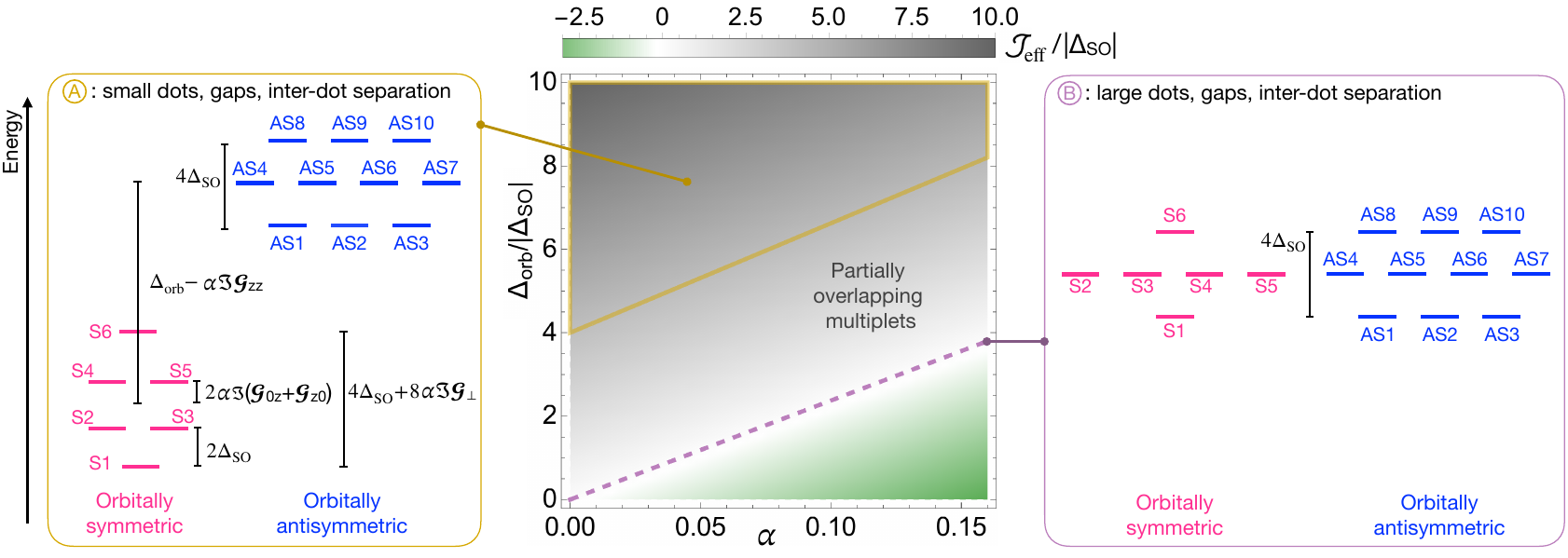}
    \caption{The ordering and multiplicity of the low-energy states of two interacting electrons in a \BLG{} \DD{} (\eqns{}\ref{eqn:StatesS} and \ref{eqn:StatesAS}) varies depending on the system parameters. The  spin and valley multiplets can be fully separated, partially overlapping, or fully overlapping, depending on the orbital splitting $\Delta_{\rm orb}$, the \SO{} gap $\Delta_{\rm SO}$, and the short-range splittings (the latter being proportional to the weight of doubly occupied orbitals, $\alpha$, cf.~\eqn{}\eqref{eqn:aalpha}). In the limiting regimes of small dots, small gaps, and small inter-dot separation (\textcolor{gold}{\raisebox{.5pt}{\textcircled{\raisebox{-.9pt} {A}}}}, left) the orbitally symmetric and antisymmetric multiplets are well separated (meaning that there is a finite gap between states $|S6\rangle$ and $|AS123\rangle$), while for large dots, large gaps, and large inter-dot distance (\textcolor{taupe}{\raisebox{.5pt}{\textcircled{\raisebox{-.9pt} {B}}}}, right) the multiplets fully collapse (here we quantify the separation  by an effective exchange constant $ \mathcal{J}_{\rm eff}$ measuring the distance between $|S4\rangle$ and $|AS7\rangle$, cf.~\eqn{}\eqref{eqn:ParameffB}). Orbitally symmetric and antisymmetric states are separated by $\Delta_{\rm orb}$. Within the multiplets, different spin and valley states are split by $\Delta_{\rm SO}$ and by the short-range interactions $\propto \alpha\mathfrak{J}\Couplg_{\perp,0z,z0}$. For these particular level orderings, we assume $\Delta_{\rm SO}<0$, $\Couplg_{\perp,0z,z0}>0$, $\Couplg_{zz}\approx 10\Couplg_{\perp}$ (see text for realistic values). 
    }
    \label{fig:2}
\end{figure*}

In the screened long-range interaction between the particles in the \QD{}s \cite{cheianovGappedBilayerGraphene2012, ganchevThreeParticleComplexesTwoDimensional2015, knotheQuartetStatesTwoelectron2020}, \eqn \eqref{eqn:MLR}, $\epsilon_0$ denotes the vacuum permittivity, $\epsilon$ is the encapsulating substrate material’s dielectric constant, and the screening length $R_{\star}=\sqrt{32}\hbar \kappa / \sqrt{m\Delta_0}$  accounts for gapped \BLG{}’s polarisability $\kappa^2= 2m e^4 / (4\pi \epsilon_0 \epsilon \hbar \sqrt{\Delta_0})^2$ with $m=\gamma_1 / 2v^2$  the effective mass \cite{cheianovGappedBilayerGraphene2012, mccannElectronicPropertiesBilayer2013} and $\Delta_0$ the \BLG{} gap.

The short-range interactions, \eqn \eqref{eqn:MSR}, are parametrized by the coupling constants, $\Couplg_{\mu\nu}$. Specifically, inter-valley scattering generates the coupling $\Couplg_{xx}=\Couplg_{yy}=\Couplg_{xy}=\Couplg_{yx}=:\Couplg_{\perp}$, intra-valley scattering leads to $\Couplg_{zz}$, and  ‘current–current’ interactions induce $\Couplg_{0z}$ and $\Couplg_{z0}$ \cite{aleinerSpontaneousSymmetryBreaking2007} (the latter favouring states with spontaneously broken time-reversal invariance \cite{lemonikCompetingNematicAntiferromagnetic2012}). These short-range coupling constants mentioned above have been shown both theoretically \cite{knotheQuartetStatesTwoelectron2020, knotheTunnelingTheoryBilayer2022} and experimentally \cite{mollerProbingTwoElectronMultiplets2021} to be crucial for providing an accurate description of the interacting few-particle confined states in \BLG{} \QD{}s\footnote{Other combinations of indices $\mu , \nu$ which do not appear equation \eqn \eqref{eqn:MSR} do not affect the states in gapped \BLG{} since the corresponding fluctuations, while allowed by the symmetry of the lattice, are suppressed by the layer polarization (Except the case $\mu = \nu = 0$ which already included in \eqn \eqref{eqn:MLR}).}.

 \emph{Single, isolated \QD{}s}, have been calculated and discussed previously in \reffs{}\onlinecite{knotheQuartetStatesTwoelectron2020, garreisShellFillingTrigonal2021, tongTunableValleySplitting2021, mollerProbingTwoElectronMultiplets2021, knotheTunnelingTheoryBilayer2022, arXiv:2305.09284, mayerTuningConfinedStates2023}. A single dot's orbital spectrum and wave functions, $\varphi_{\SPorbital}(\mathbf{r})$, are readily obtained from \eqn{}\eqref{eqn:H} with $x_l=x_r=0$ and $V_{\rm neck}\equiv0$ using the numerical diagonalisation methods of \reffs{}\onlinecite{knotheInfluenceMinivalleysBerry2018, knotheQuartetStatesTwoelectron2020, mayerTuningConfinedStates2023}. Each orbital level further splits into four spin and valley states (cf.~\fig{}\ref{fig:1}), 
\begin{equation}
E_{\SPorbital,\SPspin,\SPvalley}=E_{\SPorbital}+\SPspin\SPvalley\Delta_{\rm SO}+\SPspin \frac{1}{2} g_{S}\mu_{B} B +\SPvalley \gVSP\mu_{B} B,
\label{eqn:SPEn}
\end{equation}
where $E_{\SPorbital}$ is the \SP{} energy of the $\SPorbital$-th level,  and  $\gVSP$ is the valley g-factor. The latter is a consequence of gapped \BLG{}'s nontrivial Bloch band Berry curvature entailing an topological orbital magnetic moment  with opposite sign in the two different valleys \cite{xiaoBerryPhaseEffects2010, moulsdaleEngineeringTopologicalMagnetic2020, parkValleyFilteringDue2017, Fuchs2010}. It hence splits the valley states in a finite perpendicular magnetic field depending on the distribution of the $\SPorbital$-th level's wave function distribution in momentum space \cite{tongTunableValleySplitting2021, knotheQuartetStatesTwoelectron2020, leeTunableValleySplitting2020}.
 
 In the following, we will consider the experimentally relevant regime of weakly gapped dots for which the lowest orbital level is singly-degenerate and well separated from the higher-energy states \cite{knotheQuartetStatesTwoelectron2020}\footnote{The generalisation to considering multiple orbitals per dot is straightforward and will be explored in future work.}. We will henceforth  focus on this lowest-energy \SP{} orbital and drop the orbital index and denote the lowest-orbital \SP{}  wave function by $\varphi(\mathbf{r})$. 
 Using these orbital wave functions to write the wave functions of the left ($l$) and the right ($r$) single dots as $\varphi_{ l/r}(\mathbf{r})=\varphi (x-x_{l/r},y)\; e^{-i\frac{eB}{c\hbar}\frac{x_{l/r}}{2}}$, we obtain the orthonormalised orbital states of the \DD{} as $(\SPState_{l} , \SPState_{ r} )^{T} = \mathcal{O}^{-\frac{1}{2}} (\varphi_{l} , \varphi_{ r}  )^{T}$, where the matrix elements of the overlap matrix, $\mathbf{ \mathcal{O}}$, are given by $\mathcal{O}_{\alpha \beta}=\int d\mathbf{r}  \varphi_{ \alpha}^{*} (\mathbf{r})\varphi_{\beta}(\mathbf{r})$ with $\alpha, \beta \in \{l, r\}$.

 From these orthonormalised states, we evaluate the Hubbard parameters for \cite{burkardCoupledQuantumDots1999, morales-duranNonlocalInteractionsMoire2022, campbellModelingElectronelectronInteractions1990, gottingMoireBoseHubbardModelInterlayer2022, duttaNonstandardHubbardModels2015, jurgensenDensityinducedProcessesQuantum2012, yangGenericHubbardModel2011, leExtendedHubbardModel2017, zendra2023nonstandard}
 \begin{align}
\nonumber   &\text{the single-particle tunnelling, } t=\langle l \!:\! \SPspin \SPvalley|H_0| r\!:\! \SPspin \SPvalley \rangle,\\
\nonumber   &\text{direct nearest-neighbor interaction, } \\ \nonumber &U_{1}=\langle l\!:\!\SPspin\SPvalley,r:\SPspin^{\prime}\SPvalley^{\prime}|\VLR|l\!:\!\SPspin \SPvalley ,r\!:\!\SPspin^{\prime}\SPvalley^{\prime}\rangle,\\
\nonumber   &\text{intersite-exchange, } X=\langle l\!:\!\SPspin\SPvalley,r\!:\!\SPspin^{\prime}\SPvalley^{\prime}|\VLR|r\!:\!\SPspin \SPvalley ,l\!:\!\SPspin^{\prime}\SPvalley^{\prime}\rangle,\\
\nonumber   &\text{density-assisted hopping, } A=\langle l\!:\!\SPspin\SPvalley,l\!:\!\SPspin^{\prime}\SPvalley^{\prime}|\VLR|l\!:\!\SPspin \SPvalley ,r\!:\!\SPspin^{\prime}\SPvalley^{\prime}\rangle,\\
   &\text{pair-hopping, } P=\langle l\!:\!\SPspin\SPvalley,l\!:\!\SPspin^{\prime}\SPvalley^{\prime}|\VLR|r\!:\!\SPspin \SPvalley ,r\!:\!\SPspin^{\prime}\SPvalley^{\prime}\rangle,
   \label{eqn:Hubbards1}
 \end{align}
 and the on-site interactions which are modified by $\VSR$:
\begin{align}
 \nonumber& \langle l\!:\!\SPspin\SPvalley,l\!:\!\SPspin^{\prime}\SPvalley^{\prime}|\Hh_{C}|l\!:\!\SPspin \SPvalley ,l\!:\!\SPspin^{\prime}\SPvalley^{\prime}\rangle
= U_{0} + \mathfrak{J} [\Couplg_{zz}+\SPvalley\SPvalley^{\prime}(\Couplg_{z0}+\Couplg_{0z})],\\
 & \nonumber \langle l\!:\!\SPspin\SPvalley,l\!:\!\SPspin^{\prime}\SPvalley^{\prime}|\Hh_{C} |l\!:\!\SPspin \SPvalley^{\prime},l\!:\!\SPspin^{\prime}\SPvalley \rangle
=   4\mathfrak{J}\Couplg_{\perp},\\
  \label{eqn:Hubbards2}
\end{align}
with $ U_{0}=   \langle l\!:\!\SPspin\SPvalley,l\!:\!\SPspin^{\prime}\SPvalley^{\prime}| \VLR  |l\!:\!\SPspin \SPvalley ,l\!:\!\SPspin^{\prime}\SPvalley^{\prime}\rangle$ being the on-site Coulomb repulsion.
In above \eqn{}\eqref{eqn:Hubbards2}, we dropped the indices of $\mathfrak{J}$ in \eqn{}\eqref{eqn:MSR} as we are considering only the lowest orbital states in the left or right dot, respectively.  We relate the Hubbard parameters in \eqns{}\eqref{eqn:Hubbards1} and \eqref{eqn:Hubbards2} to the generalised Hubbard Hamiltonian of \eqn{}\eqref{eqn:HHub} in Appendix~\ref{app:RelU}.

Equipped with the Hubbard Hamiltonian, \eqn\eqref{eqn:HHub}, thus parameterised for the \BLG{} \DD{}, we study its two-electron low-energy states by exact diagonalisation using the python package qmeq \cite{kirsanskasQmeQOpensourcePython2017}. 
 In choosing realistic values for the material parameters, we base our discussion on  recent experiments \cite{mollerProbingTwoElectronMultiplets2021, knotheTunnelingTheoryBilayer2022} in  \BLG{} \QD{}s  measuring the various interaction-induced and field-induced splitting scales. For the numerical results we use $\Delta_{\rm SO}=-0.04$ meV, \cite{PhysRevB.86.245411, banszerusObservationSpinOrbitGap2020, banszerusSpinvalleyCouplingSingleelectron2021, kurzmannKondoEffectSpin2021} $\mathfrak{J}\Couplg_{\perp}\approx 0.075 $ meV, $\Couplg_{0z}\approx\Couplg_{z0}\approx\frac{2}{3}\Couplg_{\perp}$, $\Couplg_{zz}\approx 10\Couplg_{\perp}$, \cite{knotheQuartetStatesTwoelectron2020, mollerProbingTwoElectronMultiplets2021, knotheTunnelingTheoryBilayer2022}  and $\epsilon =5$  corresponding to the order of magnitude of bulk hBN widely used as encapsulating material \cite{laturiaDielectricPropertiesHexagonal2018, gottingMoireBoseHubbardModelInterlayer2022, levinshteinPropertiesAdvancedSemiconductor2001}. We provide all explicit numerical results, including other values of $\epsilon$, in  Appendix~\ref{app:ExtDat}.

Figure \ref{fig:2} illustrates the different  level orderings of \DD{} \TP{} spin and valley states we find as a function of dot and gap size and inter-dot separation. In general, the lowest-energy \TP{} states comprise a multiplet of six spin and valley states with a symmetric orbital wave function (so-called ``supersinglet" states\cite{palyiHyperfineinducedValleyMixing2009}), and ten states with an antisymmetric orbital wave function (``supertriplets" \cite{palyiHyperfineinducedValleyMixing2009}) \cite{ davidEffectiveTheoryMonolayer2018, rohlingUniversalQuantumComputing2012, peiValleySpinBlockade2012, knotheQuartetStatesTwoelectron2020, mollerProbingTwoElectronMultiplets2021, knotheTunnelingTheoryBilayer2022}. 

\begin{figure}[t]
    \centering
    \includegraphics[width=1\linewidth]{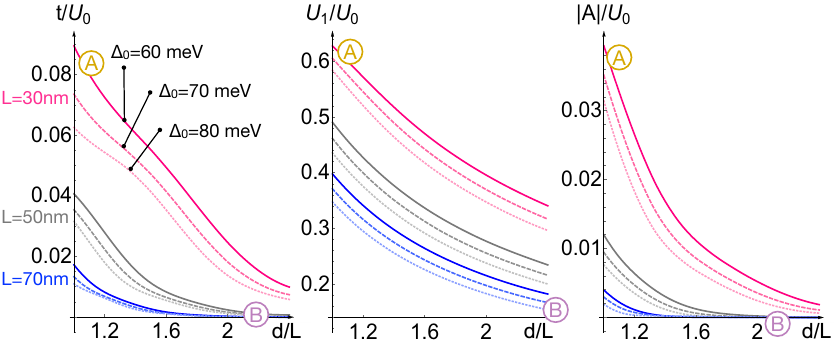}
    \caption{Hubbard parameters (\SP{} hopping, $t$, nearest neighbour direct interaction, $U_1$ and density-assisted hopping, $A$) for different dot diameters, $L$, and gaps, $\Delta_0$,  as a function of the inter-dot distance $d$. Inter-dot interaction processes are sizable compared to the on-site interaction $U_0$ over an extensive range of system parameters, making the description of \BLG{} multi-dots in terms of an extended Hubbard model necessary. The labels \textcolor{gold}{\raisebox{.5pt}{\textcircled{\raisebox{-.9pt} {A}}}} and \textcolor{taupe}{\raisebox{.5pt}{\textcircled{\raisebox{-.9pt} {B}}}} refer to the different regimes of \TP{} level schemes in \fig{}\ref{fig:2}. Here, the dots are oriented along the x-axis, we provide similar data for dots oriented along the y-axis in Appendix~\ref{app:ExtDat}.
    }
    \label{fig:4}
\end{figure}

The states with symmetric orbital wave function, $|\Psi_{S}\rangle$, are given by, 
\begin{align}
\nonumber |S1\RD =&\frac{|\Psi_{S}\rangle}{\sqrt{2}C} \Big[ \big(|+\uparrow\rangle |-\downarrow\rangle - |-\downarrow\rangle |+\uparrow\rangle \big)\\
\nonumber &+b\big( |+\downarrow\rangle |-\uparrow\rangle - |-\uparrow\rangle |+\downarrow\rangle \big)\Big],\\
\nonumber|S2\RD =& \frac{|\Psi_{S}\rangle}{\sqrt{2}}\Big[ |-\downarrow\rangle |+\downarrow\rangle - |+\downarrow\rangle |-\downarrow\rangle   \Big],\\
\nonumber|S3\RD =& \frac{|\Psi_{S}\rangle}{\sqrt{2}}  \Big[  |+\uparrow\rangle |-\uparrow\rangle - |-\uparrow\rangle |+\uparrow\rangle  \Big] ,\\
\nonumber|S4\RD = & \frac{|\Psi_{S}\rangle}{\sqrt{2}}   \Big[ |+\downarrow\rangle |+\uparrow\rangle - |+\uparrow\rangle |+\downarrow\rangle   \Big],\\
\nonumber|S5\RD =&  \frac{|\Psi_{S}\rangle}{\sqrt{2}}   \Big[ |-\uparrow\rangle |-\downarrow\rangle - |-\downarrow\rangle |-\uparrow\rangle   \Big] ,\\
\nonumber|S6\RD =&\frac{|\Psi_{S} \rangle}{\sqrt{2}C}\Big[\big( |+\downarrow\rangle |-\uparrow\rangle - |-\uparrow\rangle |+\downarrow\rangle \big) \\
&-b  \big(|+\uparrow\rangle |-\downarrow\rangle - |-\downarrow\rangle |+\uparrow\rangle \big) \Big] ,
\label{eqn:StatesS}
\end{align}

with $|\SPspin\SPvalley\rangle$ denoting the spin ($\SPspin=\uparrow,\downarrow$) and valley ($\SPvalley=+,-$) state and \cite{burkardCoupledQuantumDots1999}\footnote{To arrive at this simple analytical expression, we assume $A=X=P\approx0$, which is justified for the system under consideration, see \fig{}\ref{fig:4}, and describes the numerical data well.},
\begin{equation}
    |\Psi_{S}\rangle\approx\Big[\frac{a_1}{\sqrt{2}} 
\big(|l\rangle|r\rangle+|r\rangle|l\rangle \big)+a_2\big( |l\rangle|l\rangle+|r\rangle|r\rangle \big)\Big].
\end{equation}
The orbital coefficients, $a_1$ and $a_2$, are given by,
\begin{align}
 \nonumber   a_1=&  \frac{1}{\sqrt{1+\frac{16t^2}{(U_0-U_1+\sqrt{16t^2+(U_0-U_1)^2})^2}}},\\
\nonumber a_2= &\frac{1}{\sqrt{2}}\sqrt{\frac{1+\frac{16t^2}{(U_0-U_1+\sqrt{16t^2+(U_0-U_1)^2})^2 }}{4+\frac{(U_0-U_1)^2}{4t^2}}}\approx\frac{1}{\sqrt{2}}\sqrt{\alpha},\\
&\text{ where } \alpha\approx\frac{1}{3+\frac{(U_0-U_1)^2}{4t^2}} \text{ for } \frac{4t}{(U_0-U_1)}\ll1,
\label{eqn:aalpha}
\end{align}
quantifying the occupation of the left/right symmetrised orbitals and of the doubly occupied orbitals, respectively. Conversely, the coefficient\footnote{This expression \eqn{}\ref{eqn:b} has been derived in first order perturbation theory for $\hat\Upsilon$ treating the short-range interactions as a perturbation and assuming non-degenerate levels. This expression hence only holds for $\Delta_{\rm SO} \neq 0$. }
\begin{equation}
   b=\frac{\alpha \Couplg_{\perp}\mathfrak{J}}{\Delta_{\rm SO} }, 
   \label{eqn:b}
\end{equation}
quantifies the admixture with higher energy spin and valley states induced by the short-range inter-valley scattering. This short-range interaction $\propto\Couplg_{\perp}$ couples different valleys when electrons doubly occupy a site ($\propto \alpha$) for states of the same spin (states $|S_1\rangle$ and $|S_6\rangle$). Note that all other short-range interaction contributions do not couple different valley states and therefore only lead to shifts in energy $\propto \Couplg_{zz, 0z,z0}$ without inducing mixing between different states. Lastly,
\begin{equation}
C = \sqrt{1+b^2},
\end{equation}
ensures normalisation.

The states with the orbitally antisymmetric wave function \cite{burkardCoupledQuantumDots1999}, 
\begin{equation}
     |\Psi_{AS}\rangle=  \frac{1}{\sqrt{2}}\big[
|l\rangle|r\rangle-|r\rangle|l\rangle \big],
\end{equation}
comprise the multiplet, 
\begin{align}
\nonumber |AS1\RD &= \frac{|\Psi_{AS}\rangle}{\sqrt{2}}\big[ |-\downarrow\rangle |+\uparrow\rangle + |+\uparrow\rangle |-\downarrow\rangle   \big] ,\\
\nonumber|AS2\RD &= |\Psi_{AS}\rangle  |-\downarrow\rangle |-\downarrow\rangle   ,\\
\nonumber|AS3\RD &= |\Psi_{AS}\rangle  |+\uparrow\rangle |+\uparrow\rangle    ,\\
\nonumber|AS4\RD &= \frac{|\Psi_{AS}\rangle}{\sqrt{2}}\big[ |-\downarrow\rangle |-\uparrow\rangle + |-\uparrow\rangle |-\downarrow\rangle    \big] ,\\
\nonumber|AS5\RD &=\frac{|\Psi_{AS}\rangle}{\sqrt{2}} \big[   |-\uparrow\rangle |+\uparrow\rangle + |+\uparrow\rangle |-\uparrow\rangle  \big] ,\\
\nonumber|AS6\RD &=\frac{|\Psi_{AS}\rangle}{\sqrt{2}}\big[ |+\downarrow\rangle |-\downarrow\rangle + |-\downarrow\rangle |+\downarrow\rangle  \big] ,\\
\nonumber|AS7\RD &= \frac{|\Psi_{AS}\rangle}{\sqrt{2}}\big[ |+\uparrow\rangle |+\downarrow\rangle + |+\downarrow\rangle |+\uparrow\rangle  \big] ,\\
\nonumber|AS8\RD &=  \frac{|\Psi_{AS}\rangle}{\sqrt{2}}\big[ -\uparrow\rangle |+\downarrow\rangle + |+\downarrow\rangle |-\uparrow\rangle   \big],\\
\nonumber|AS9\RD &= |\Psi_{AS}\rangle  |-\uparrow\rangle |-\uparrow\rangle    ,\\
|AS10\RD &=  |\Psi_{AS}\rangle  |+\downarrow\rangle |+\downarrow\rangle      .
\label{eqn:StatesAS}
\end{align}
We relate the orbitally symmetric and antisymmetric spin and valley multiplets to product states of spin/valley singlet and triplet states in Appendix~\ref{app:RelST}.
\begin{figure}[b]
    \centering
    \includegraphics[width=1\linewidth]{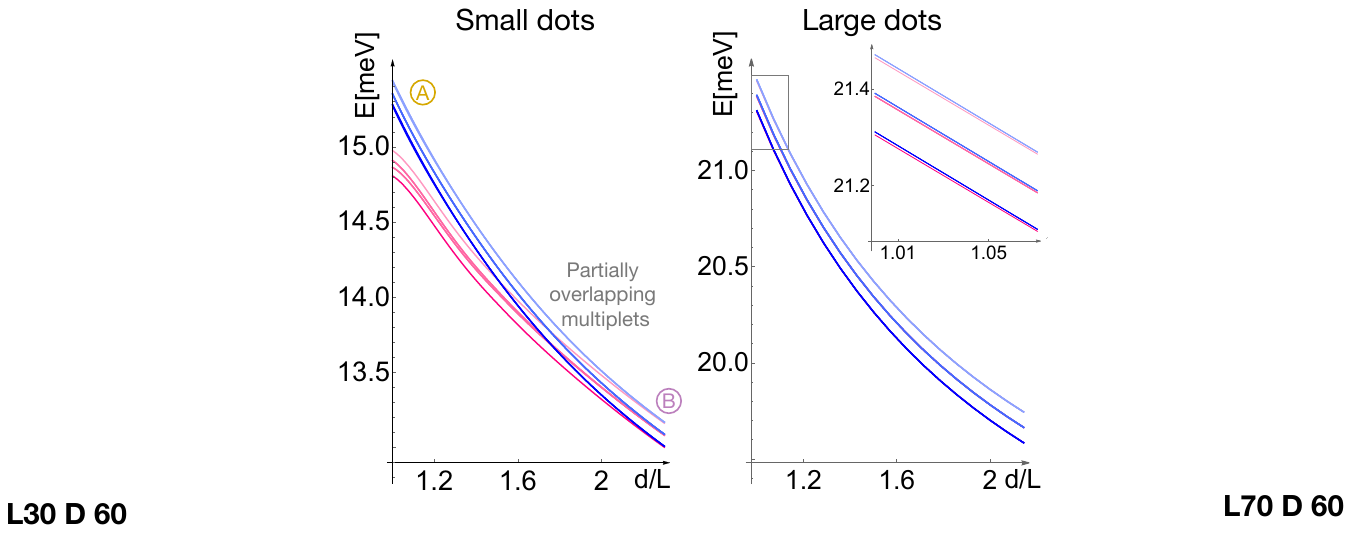}
    \caption{The two-particle spectra for small dots and small gaps (left, $L=30$ nm, $\Delta_0=60$ meV) and large dots and gaps (right, $L=70$ nm, $\Delta_0=80$ meV) as a function of inter-dot separation demonstrate the different regimes for the state orderings,  \textcolor{gold}{\raisebox{.5pt}{\textcircled{\raisebox{-.9pt} {A}}}} and  \textcolor{taupe}{\raisebox{.5pt}{\textcircled{\raisebox{-.9pt} {B}}}} and the state splittings as discussed in \fig{}\ref{fig:2}. Here, the dots are oriented along the x-axis and we assume realistic values for the couplings and the spin-orbit splittings, $\mathfrak{J}\Couplg_{\perp}\approx 0.075 $ meV, $\Couplg_{0z}\approx\Couplg_{z0}\approx\frac{2}{3}\Couplg_{\perp}$, $\Couplg_{zz}\approx 10\Couplg_{\perp}$, and $\Delta_{\rm SO}\approx - 0.04$ meV. {We provide spectra for different interaction strengths in Appendix \ref{app:ExtDat}.}
    }
    \label{fig:3}
\end{figure}

We demonstrate the regimes of possible level orderings in \fig\ref{fig:2} and provide explicit numerical results in for their dependence on dot size, gap size, and interdot-distance in representative parameter regimes in \fig{}\ref{fig:3}. 
The states with symmetric orbital wave function, \eqn{}\eqref{eqn:StatesS}, generally yield the lowest-energy states, separated from orbitally antisymmetric states of \eqn{}\eqref{eqn:StatesAS} by a splitting $\sim \Delta_{\rm orb}-\alpha\mathfrak{J}\Couplg_{zz}$, 
dominated by the orbital splitting\cite{burkardCoupledQuantumDots1999} 
\begin{equation}
\Delta_{\rm orb}\approx-\frac{U_0-U_1}{2}+\frac{1}{2}\sqrt{(U_0-U_1)^2+16t^2}.
\label{eqn:Deltaorb}
\end{equation}
Within each multiplet, the \SO{} splitting, $\Delta_{\rm SO}$, dominates the splittings between different spin and valley states. Additionally, due to the finite weight of doubly occupied sites for states with a symmetric orbital wave function, cf.~\eqn \eqref{eqn:StatesS}, the short-range interactions of \eqn{}\eqref{eqn:MSR} induce an additional shift ($\propto\Couplg_{zz}$) and splittings ($\propto\Couplg_{0z}, \Couplg_{z0},\Couplg_{\perp}$) of the orbitally symmetric multiplet. These   short-range-induced shifts are proportional to the fraction of double occupation, $\alpha$, and the short-range coupling constants, $\Couplg_{\perp, zz, 0z,z0} $.

The smaller the dots, the gaps, and the inter-dot distances, the larger the  orbital splittings, \eqn{}\eqref{eqn:Deltaorb}, and the short-range induced splittings. Hence, in this regime of small and nearby dots (\textcolor{gold}{\raisebox{.5pt}{\textcircled{\raisebox{-.9pt} {A}}}} in \figs{}\ref{fig:2} and \ref{fig:3}), the multiplets of different orbital symmetry are  well separated and the ground state is given by an orbitally symmetric state with vanishing total spin and valley pseudospin uniquely selected by the \SO{} splitting (state $|S1\rangle$ in \eqn{}\eqref{eqn:StatesS}, cf.~\fig{}\ref{fig:2}). In the opposite limit, for large dots, gaps, and separation between the dots (\textcolor{taupe}{\raisebox{.5pt}{\textcircled{\raisebox{-.9pt} {B}}}} in \figs{}\ref{fig:2} and \ref{fig:3}), the  orbital splitting, \eqn{}\eqref{eqn:Deltaorb}, and the short-range induced splittings ($\propto\alpha$ in \eqn{}\eqref{eqn:aalpha}) vanish and the multiplets collapse such that the ground state is four-fold degenerate with orbitally symmetric and antisymmetric states at the same energy, see \figs{}\ref{fig:2} and \ref{fig:3}. 

We demonstrate these different regimes of separated (\textcolor{gold}{\raisebox{.5pt}{\textcircled{\raisebox{-.9pt} {A}}}}) or collapsed (\textcolor{taupe}{\raisebox{.5pt}{\textcircled{\raisebox{-.9pt} {B}}}}) multiplets induced by the interplay of the splittings in \fig\ref{fig:3}, where we compare numerical spectra for small \DD{}s and for large \DD{}s  as a function of the inter-dot distance, $d$.

We can relate the parameter dependence of the orbital splitting, $\Delta_{\rm orb}$ in \eqn{}\eqref{eqn:Deltaorb}, and the weight of double occupation, $\alpha$ in \eqn{}\eqref{eqn:aalpha}, (determining the short-range splittings) to the Hubbard parameters, \eqns{}\eqref{eqn:Hubbards1} and \eqref{eqn:Hubbards2}. Figure \ref{fig:4} exemplifies the dominant Hubbard parameters, $t, U_1$, and $A$ for different dot and gap sizes and dot orientations. We find non-local  parameters of an extended Hubbard model to be sizable over a large range pf system parameters: For small dots,  gaps, and  inter-dot distances both the hopping and the extended Hubbard parameters manifest compared to the onsite Hubbard $U_0$. The fast decay of the hopping and the density induced tunnelling with $d$ leaves  the direct interaction as the dominant non-local Hubbard parameter at larger inter-dot distances. Conversely, we found the exchange parameters, $X$, and the pair-wise hopping parameters, $P$, to be negligibly small compared to the other Hubbard terms (we show the data for all parameters explicitly in the Appendix \ref{app:ExtDat}).

\section{Double-dot in a magnetic field: Effective Spin Heisenberg model}

\begin{figure}[b]
    \centering
    \includegraphics[width=1\linewidth]{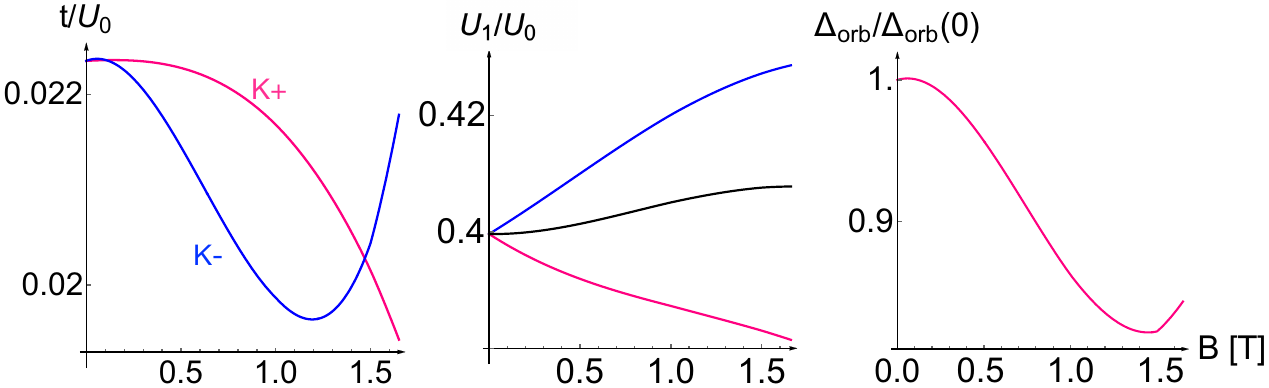}
    \caption{Extended Hubbard parameters (tunnelling, $t$, and nearest-neighbor direct interaction $U_1$) and the orbital splitting, $\Delta_{\rm orb}$, as a function of magnetic field, B, for  $L=50$ nm,  $\Delta_0=70$ meV, and $d=1.2L$ for two dots aligned along the x-axis. Magenta/blue/black lines represent processes involving states in the   $K^{+}$/$K^{-}$/ either valley. Changes in the orbital splitting due to these field-induced modulations of the Hubbard parameters are weak compared to the overall splitting scale (rightmost panel). We provide the corresponding data for dots oriented along the y-axis in Appendix~\ref{app:ExtDat}
    }
    \label{fig:5}
\end{figure}

A perpendicular magnetic field splits the spin and the valley states, respectively, according to their g-factors, $g_s$ and $g_v$. Hence, at finite B, the \SP{} wave function changes due to this shift in energy within a soft confinement potential whose radius changes with energy, the compression by the magnetic field, and the B-dependent phase factor in the \SP{} wave functions $\varphi_{ l/r}(\mathbf{r})$ of the individual dots. These B-dependent effects combined entail a non-monotonic dependence of the Hubbard parameters on the magnetic field strength as we demonstrate in \fig{}\ref{fig:5}.

 We find that these orbital magnetic field effects are small compared to the field induced valley splitting, see \fig{}\ref{fig:5}, right panel. 
The valley g-factor (induced by the topological orbital magnetic moment of the \BLG{} Bloch bands)  can be orders of magnitude larger than the spin g-factor \cite{knotheQuartetStatesTwoelectron2020, mayerTuningConfinedStates2023, leeTunableValleySplitting2020, mollerProbingTwoElectronMultiplets2021, moulsdaleEngineeringTopologicalMagnetic2020, overwegTopologicallyNontrivialValley2018, tongTunableValleySplitting2021} and hence valley Zeeman splitting   dominates the \SP{} level ordering at finite magnetic field.

Due to this prevalence of the field-induced valley splitting, the \DD{}'s low-energy multiplet in a finite magnetic field consists of the  valley polarized \TP{} states. By projecting onto the well-separated valley polarized states, we  describe the low-energy multiplet by an effective Heisenberg model for the spins, \cite{readFeaturesPhaseDiagram1989},
\begin{equation}
    \Hh_{\rm eff,B}= E_{V}+ \mathcal{J}_{\rm eff} \mathbf{S}_l\cdot\mathbf{S}_r + \mu_B\mathcal{B}_{\rm eff}(S^z_{l}+S^z_{r}),
     \label{eqn:HeffB}
\end{equation}
where 
\begin{equation}
     E_{V}= g_v\mu_B B,
\end{equation}
is the global energy of the valley polarised multiplet and,
\begin{align}
  \nonumber  \mathcal{J}_{\rm eff}&= \Delta_{\rm orb}-\alpha 2 \mathfrak{J}(\Couplg_{0z}+\Couplg_{z0}+\frac{1}{2}\Couplg_{zz}),  \\
   \mathcal{B}_{\rm eff}&=2\frac{\Delta_{\rm SO}}{\mu_B}+g_s B,
   \label{eqn:ParameffB}
\end{align}
represent an effective exchange coupling and an effective magnetic field capturing the combined effect of the external magnetic field, $B$, and the material's characteristics.

 \begin{figure}[htb]
    \centering
    \includegraphics[width=0.9\linewidth]{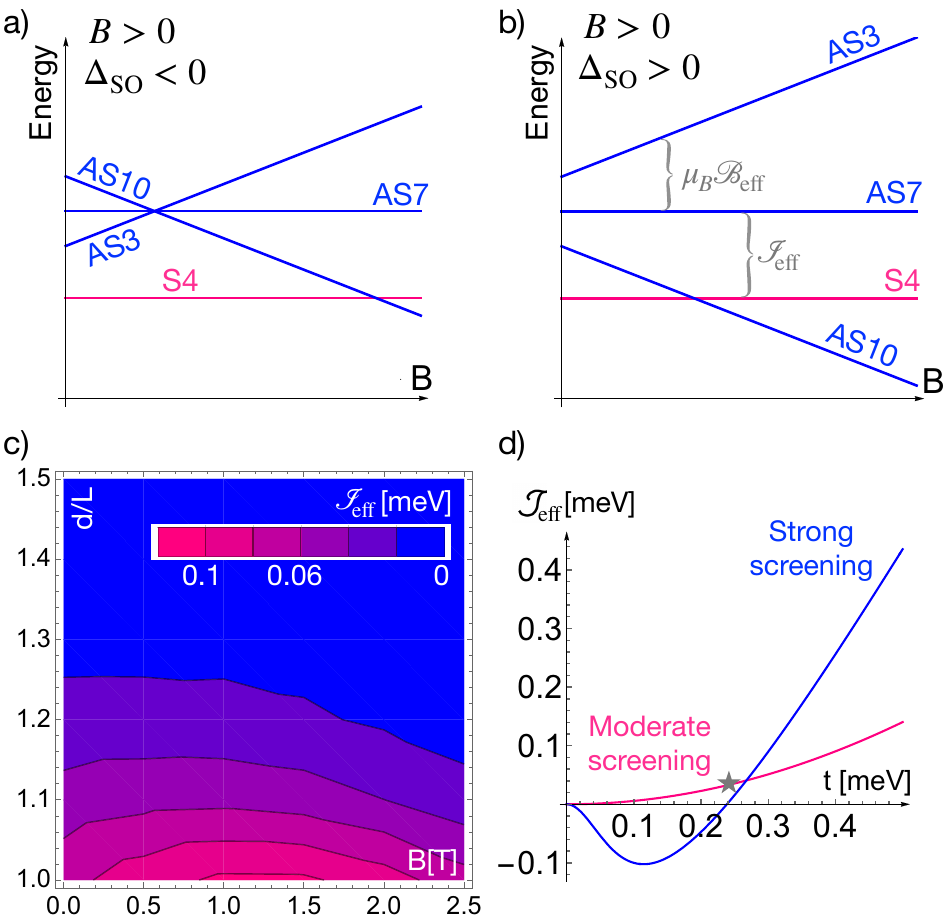}
    \caption{Top: Level orderings in the low-energy valley polarized multiplet at finite magnetic field $B>0$ described by the effective anisotropic Heisenberg model, \eqn{}\eqref{eqn:HeffB} for different   sign of the \SO{}-splitting, $\Delta_{\rm SO}<0$ (a) or $\Delta_{\rm SO}>0$ (b). For clarity, we omitted the global energy shift, $E_V$, of the valley polarized multiplet and the explicit dependence of  $ \mathcal{J}_{\rm eff}$ on $B$.  Bottom: Effective exchange constant, $ \mathcal{J}_{\rm eff}$, as a function of parameters. c) $ \mathcal{J}_{\rm eff}$ as a function of magnetic field $B$ and inter-dot distance $d$ evaluated from the data in \fig{}\ref{fig:5}. d) $ \mathcal{J}_{\rm eff}$ as a function of the hopping $t$ assuming the screening of extended Hubbard parameters to be of different strengths. The magenta line is obtained using $U_0$ and $U_1$ for the parameters in \fig{}\ref{fig:5} (the star marking the corresponding value of $t$). The blue line assumes $U_0$ and $U_1$ to be reduced by a factor of 10, allowing for negative values as the orbital interactions become comparable to the short-range interactions.
    }
    \label{fig:6}
\end{figure}

We examine the effective Heisenberg model of \eqn{}\eqref{eqn:HeffB} describing the four valley polarized low-energy spin singlet and triplet states  in \fig{}\ref{fig:6}. 
In the effective magnetic field, $\mathcal{B}_{\rm eff}$, \eqn{}\eqref{eqn:ParameffB},  the Zeeman coupling and the \SO{} gap compete. This competition entails different level orderings depending on the signs and strengths of $\Delta_{\rm SO}$ and $\mu_B B$, as demonstrated in the top panels of  \fig{}\ref{fig:6} for a positive magnetic field which singles out the $K^+$ polarized multiplet in our convention. The  \SO{} and the Zeeman effect favouring different spin and valley states (here, this is the case for $\Delta_{\rm SO}<0$), leads to a reversal of the order within the orbitally antisymmetric states at a critical field strength, see \fig{}\ref{fig:6} a). Conversely, when $\Delta_{\rm SO}$ and $\mu_B B$  favour the same spin and valley state (here, $\Delta_{\rm SO}>0$), no such state reordering occurs, c.f.~\fig{}\ref{fig:6} b). For sufficiently large field strengths, the  valley and spin-polarized state favoured by the Zeeman coupling becomes the global \GS{} in either case. A negative magnetic field entails the corresponding level schemes in the $K^-$ polarized multiplet. We show the  orderings of all the states in the Appendix \ref{app:B}.

For the effective exchange constant, $\mathcal{J}_{\rm eff}$, \eqn{}\eqref{eqn:ParameffB}, the  contributions from orbital and short-range interactions compete. 
A dominant orbital splitting, $\Delta_{\rm orb}$, entails $\mathcal{J}_{\rm eff}>0$. Conversely, sufficiently strong screening of the orbital contributions, such that $|U_0 - U_1|< 2 \mathfrak{J}(\Couplg_{0z}+\Couplg_{z0}+\frac{1}{2}\Couplg_{zz})$, allows for negative values of $\mathcal{J}_{\rm eff}$ for small hoppings, $t$, as we demonstrate in the right panel of \fig{}\ref{fig:6}. Being able to control the effective exchange constant would be useful in a scenario using such \DD{}s, e.g., as 2-qubit gates\cite{burkardCoupledQuantumDots1999,Martins2017,Deng2018}. Further, the possibility to control  $\mathcal{J}_{\rm eff}$ via multiple parameters (such as screening, inter-dot distance, magnetic field, and the gates) will provide enhanced sensitivity against noise \cite{Culcer2009,Burkard2023}. We discuss possible ways to tune the different parameters experimentally in the conclusion.

 \section{Conclusions and Outlook}

We set up and parametrised a Hubbard model for interacting \QD{}s in \BLG{} and studied a double \QD{} as the smallest possible example. We characterised the low-energy \TP{} multiplets of the \DD{} in terms of their orbital, spin, and valley configuration and study their dependence on the system parameters and an external magnetic field. The various spin and valley phases are driven by the interplay of extended Hubbard parameters induced by long-range Coulomb interaction and short-range interactions on the lattice scale. 

 \begin{figure}[h]
    \centering
    \includegraphics[width=0.6\linewidth]{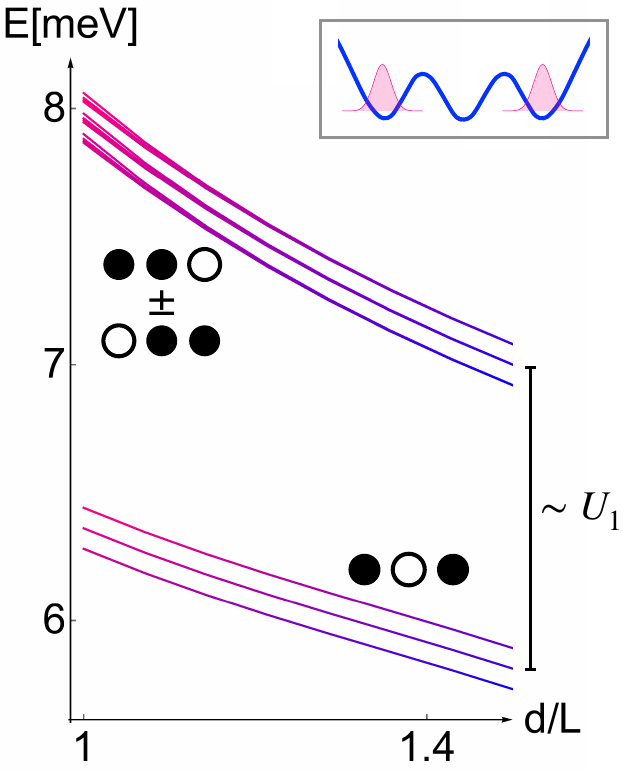}
    \caption{Spectrum of two interacting electrons in a triple dot chain oriented along the x-axis, described by a Hubbard model with the extended Hubbard parameters in \fig{}\ref{fig:4} for $L=70$ nm and $\Delta=60$ meV, plotted as a function of the inter-dot distance $d$. The \GS{} multiplet consists of the same spin and valley states as the \DD{}, where the electrons prefer to sit in the outermost sites in order to minimise their energy. Only the excited states' multiplet allows for two electrons to reside in adjacent dots. The separation between the ground and excited states' multiplets is of the order of the nearest-neighbour direct interaction, $U_1$.}
    \label{fig:TripleQD}
\end{figure}

This competition of interactions on different scales opens ample ways to manipulate and tune the dot states and the couplings. We discuss the dependence of the states and splittings on the dot size and separation, affecting the orbital wave function and the tunnelling, cf.~\figs{}\ref{fig:3}, \ref{fig:4}, and \ref{fig:6}. 

In future work, one may explore, e.g., dot state tuning by deformation of the \QD{}s into ellipses and by an independently adjustable tunnelling barrier (the former can be achieved by having split gates and finger gates of different dimensions, the latter by adding an additional finger gate in between the two \QD{}s) \cite{banszerusTunableInterdotCoupling2021, banszerusElectronHoleCrossover2020, arXiv:2211.04882, arXiv:2305.03479}. Further, we expect the long-range interactions to be strongly affected by environmental screening as opposed to short-range interactions largely confined to the \BLG{} lattice. This difference in screening response may allow for efficient dielectric engineering of the different interaction parameters \cite{PhysRevB.100.161102, vanloonCoulombEngineeringTwodimensional2023, steinleitnerDielectricEngineeringElectronic2018}

We note that the non-local Hubbard parameters for the \BLG{} \DD{} (the nearest-neighbour direct interaction, $U_1$, in particular) remain finite over an extensive range of system parameters, including the inter-dot separation, cf.~\fig{}\ref{fig:4}. Therefore, one indeed requires an extended Hubbard model framework to describe a \BLG{} multi-dot system faithfully. Long-range extended Hubbard parameters will also affect lattices with more than two dots over several dot sites, as we demonstrate for a triple \QD{} in \fig{}\ref{fig:TripleQD}. Here, the \TP{} \GS{} preferring to maximise the distance between the electrons is driven by inter-dot interactions, in particular the nearest-neighbor direct interaction $U_1$. {In future work, one may employ an extended Hubbard model approach as developed in this work, to study \BLG{} \QD{} lattices with larger numbers of dots, where additional intricacies, such as long-range inter-site interactions, band overlap, and multi-orbital physics may make for interesting Fermi-Hubbard physics.}

Besides the use of the spin and valley degree of freedom for quantum information processing in multi-qubit systems,  larger dot lattices with tunable long-range and spin and valley-dependent short-range interactions may hence also allow for quantum simulation of exotic Hubbard models using \BLG{} \QD{} lattices.

\emph{Acknowledgements.} 
We thank Samuel M\"oller, Katrin Hecker, Christoph Stampfer, Thierry Jolic\oe ur, Oded Zilberberg, Jonathan Brugger, and Dennis Mayer for fruitful discussions. AK acknowledges support from the Deutsche Forschungsgemeinschaft (DFG, German Research Foundation) within Project-ID 314695032 -- SFB 1277 and DFG Individual grant KN 1383/4.

\appendix

\section{Relation of the Hubbard parameters}
\label{app:RelU}
We label the four states in each dot with indices $\{h,j,k,m\}\in\{0,\dots,7\}$, relating $|0\rangle=|l\!:\uparrow +\rangle$, $|1\rangle= |l\!:\downarrow +\rangle$, $|2\rangle=|l\!:\uparrow -\rangle$, $|3\rangle=|l\!:\downarrow -\rangle$, $|4\rangle=|r\!:\uparrow +\rangle$ , $|5\rangle=|r\!:\downarrow +\rangle$, $|6\rangle=|r\!:\uparrow - \rangle$, $|7\rangle=|r\!:\downarrow -\rangle$. In this notation, we relate the Hubbard parameters as follows (listing the non-zero matrix elements $t_{jk}$ with $k>j$ and $U_{hjkm}$ with $j>h$ and $k>m$):
\begin{align}
\nonumber &t_{04} = t_{15} =t_{26} =t_{37} =t,\\
\nonumber  &  U_{0110}= U_{2332} = U_0 +  \mathfrak{J}[\Couplg_{zz}+\Couplg_{z0} +\Couplg_{0z}],\\
\nonumber  & U_{0220} = U_{0330} =U_{1221} = U_{1331} = U_0 +  \mathfrak{J}[\Couplg_{zz}-(\Couplg_{z0} +\Couplg_{0z})],\\
\nonumber & U_{0202} = U_{0312} = U_{1313} = U_{1203} =4 \mathfrak{J}\Couplg_{\perp},\\
\nonumber & U_{0440} = U_{0550} = U_{0660} = U_{0770} = U_{1441} = U_{1551} = U_{1661} \\
\nonumber &= U_{1771} = U_{2442} = U_{2552} = U_{2662} = U_{2772} = U_{3443}\\
\nonumber & = U_{3553} = U_{3663} = U_{3773} = U_1,\\
\nonumber & U_{0404} = U_{1515} = U_{2626} = U_{3737} = X,\\
\nonumber & U_{0154} = U_{0264} = U_{0374} = U_{1265} = U_{1375} = U_{2376} = P,\\
\nonumber & U_{1261} = U_{1371} + U_{2372} = U_{1621} = U_{0150} = U_{0260} = U_{0370} \\
\nonumber & = U_{0510} = U_{0620} = U_{0730} = U_{1401} = U_{1731} = U_{2732}\\
 &= U_{2402} = U_{2512} =U_{3403} = U_{3623} =U_{3513} = A.
\end{align}

\section{Relation to spin and valley triplet states}
\label{app:RelST}
We relate the \TP{} states in \eqns{}\eqref{eqn:StatesS} and \eqref{eqn:StatesAS} to the the spin and valley triplet states. For this appendix, we adopt notation commonly used in the literature \cite{davidEffectiveTheoryMonolayer2018} to denote the spin (s) and valley (v) singlet ($|S\rangle$) and triplet ($|T_{-,0,+}\rangle$) states as,
\begin{align}
\nonumber  &  |S\rangle^{s} = \frac{1}{\sqrt{2}} (|\uparrow\downarrow\rangle-|\downarrow\uparrow\rangle),\\
\nonumber & |T_{-}\rangle^{s} = |\downarrow\downarrow\rangle,\\
\nonumber & |T_{0}\rangle^{s} =\frac{1}{\sqrt{2}} ( |\uparrow\downarrow\rangle + |\downarrow\uparrow\rangle),\\
\nonumber & |T_{+}\rangle^{s} = |\uparrow\uparrow\rangle,\\
\nonumber  &  |S\rangle^{v} = \frac{1}{\sqrt{2}} (|+ -\rangle-|-+\rangle),\\
\nonumber & |T_{-}\rangle^{v} = |--\rangle,\\
\nonumber & |T_{0}\rangle^{v} =\frac{1}{\sqrt{2}} ( |+-\rangle + |-+\rangle),\\
\ & |T_{+}\rangle^{v} = |++\rangle.
\end{align}
and combinations thereof.

Using the notation above, we relate
\begin{align}
\nonumber  |S1\rangle &= \frac{|\Psi_{S}\rangle}{ C} \big[(|S\rangle^{v}|T_{0}\rangle^s + |T_{0}\rangle^v |S\rangle^{s})\\
\nonumber&+b(|S\rangle^{v}|T_{0}\rangle^s - |T_{0}\rangle^v |S\rangle^{s})\big],\\
\nonumber |S2\rangle &= -|\Psi_{S}\rangle |S\rangle^{v} |T_{-}\rangle^{s},\\
\nonumber |S3\rangle &= |\Psi_{S}\rangle|S\rangle^{v} |T_{+}\rangle^{s},\\
\nonumber |S4\rangle &= -|\Psi_{S}\rangle |T_+\rangle^{v} |S\rangle^{s},\\
\nonumber |S5\rangle &= |\Psi_{S}\rangle |T_-\rangle^{v} |S\rangle^{s},\\
\nonumber   |S6\rangle &= \frac{|\Psi_{S}\rangle}{ C} \big[(|S\rangle^{v}|T_{0}\rangle^s - |T_{0}\rangle^v |S\rangle^{s})\\
&+b(|S\rangle^{v}|T_{0}\rangle^s + |T_{0}\rangle^v |S\rangle^{s})\big],
\end{align}
and
\begin{align}
   \nonumber  |AS1\rangle &=    |\Psi_{AS}\rangle (|T_{0}\rangle^{v}|T_{0}\rangle^s + |S\rangle^v |S\rangle^{s}   )      ,\\
    \nonumber  |AS2\rangle &=   |\Psi_{AS}\rangle |T_{-}\rangle^{v} |T_{-}\rangle^{s}         ,\\
     \nonumber  |AS3\rangle &= |\Psi_{AS}\rangle |T_{+}\rangle^{v} |T_{+}\rangle^{s}           ,\\
      \nonumber  |AS4\rangle &=   |\Psi_{AS}\rangle |T_{-}\rangle^{v} |T_{0}\rangle^{s}           ,\\
       \nonumber  |AS5\rangle &=   |\Psi_{AS}\rangle |T_{0}\rangle^{v} |T_{+}\rangle^{s}           ,\\
 \nonumber  |AS6\rangle &=   |\Psi_{AS}\rangle |T_{0}\rangle^{v} |T_{-}\rangle^{s}           ,\\
  \nonumber  |AS7\rangle &=  |\Psi_{AS}\rangle |T_{+}\rangle^{v} |T_{0}\rangle^{s}            ,\\
   \nonumber  |AS8\rangle &=      |\Psi_{AS} \rangle(|T_{0}\rangle^{v}|T_{0}\rangle^s - |S\rangle^v |S\rangle^{s}     )  ,\\
    \nonumber  |AS9\rangle &=  |\Psi_{AS}\rangle |T_{-}\rangle^{v} |T_{+}\rangle^{s}            ,\\
       |AS10\rangle &=    |\Psi_{AS}\rangle |T_{+}\rangle^{v} |T_{-}\rangle^{s}          .
\end{align}

\section{Extended Data for the Hubbard parameters}
\label{app:ExtDat}

In this appendix, we provide numerical data for the Hubbard parameters in \eqns{}\ref{eqn:Hubbards1} and \ref{eqn:Hubbards2} {as well as for the \TP{} spectra} over a larger parameter range compared to the main text. 

\begin{figure}[h]
    \centering
    \includegraphics[width=1\linewidth]{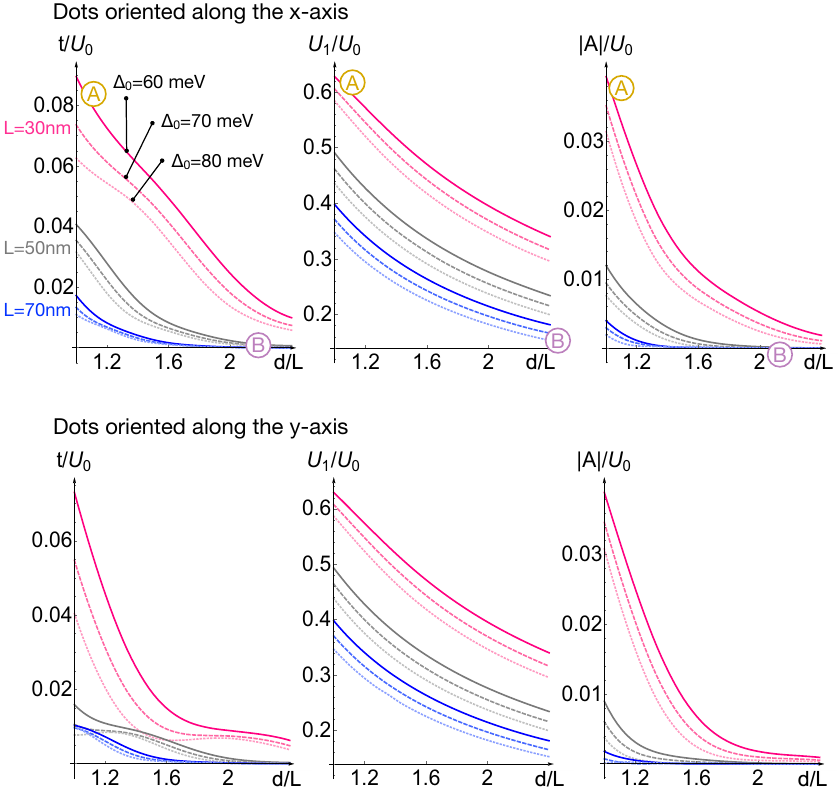}
    \caption{Hubbard parameters as a function of the inter-dot separation, $d$, for dots oriented along the $x$-axis (top) or along the $y$-axis (bottom). There is little difference for the extended Hubbard parameters, $U_1$ and $A$, while the \SP{} hopping is slightly reduced for dots aligned along the  $y$-axis (cf.~\fig{}\ref{fig:4} in the main text).
    }
    \label{fig:CompXY}
\end{figure}
Figure \ref{fig:CompXY} compares the hopping, $t$, next neighbour direct interaction, $U_1$, and the density assisted hopping, $A$ for the scenarios where the two dots are aligned along the $x$-axis, or the $y$-axis. The breaking of rotational symmetry in \BLG{} translates into the \SP{} wave functions and hence the Hubbard parameters. Nevertheless, we observe in \fig{}\ref{fig:CompXY} that the parameters are of similar order of magnitude for both orientations ( mostly the hopping, $t$, is slightly reduced for orientations along the $y$-axis as compared to the $x$-axis). We hence conclude all qualitative statements from the main text to be valid independent of the orientation of the \QD{} chain with respect to the lattice.
 \begin{figure*}[h!]
    \centering
    \includegraphics[width=0.95\linewidth]{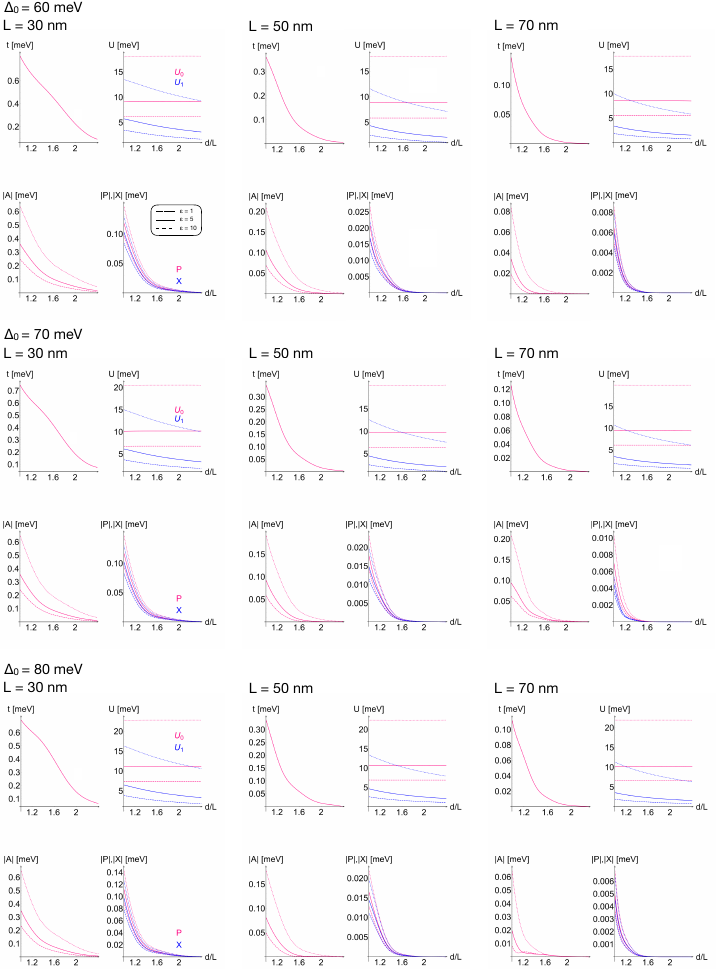}
    \caption{Hubbard parameters $t$, $U_0$, $U_1$, $P$, $A$, and $X$ (cf.~\eqns{}\ref{eqn:Hubbards1} and \ref{eqn:Hubbards2}) as a function of the inter-dot separation, $d$, over a large range of system parameters.
    }
    \label{fig:AppD607080}
\end{figure*}

Further, \fig{}\ref{fig:AppD607080} provides the unnormalised Hubbard parameters for two dots oriented along the $x$-axis over an extensive parameter range. We vary the gap, $\Delta_0$, the dot diameter, $L$, the inter-dot separation, $d$, and the dielectric constant, $\epsilon$, of the encapsulating medium. 

\begin{figure}[hb]
    \centering
    \includegraphics[width=0.99\linewidth]{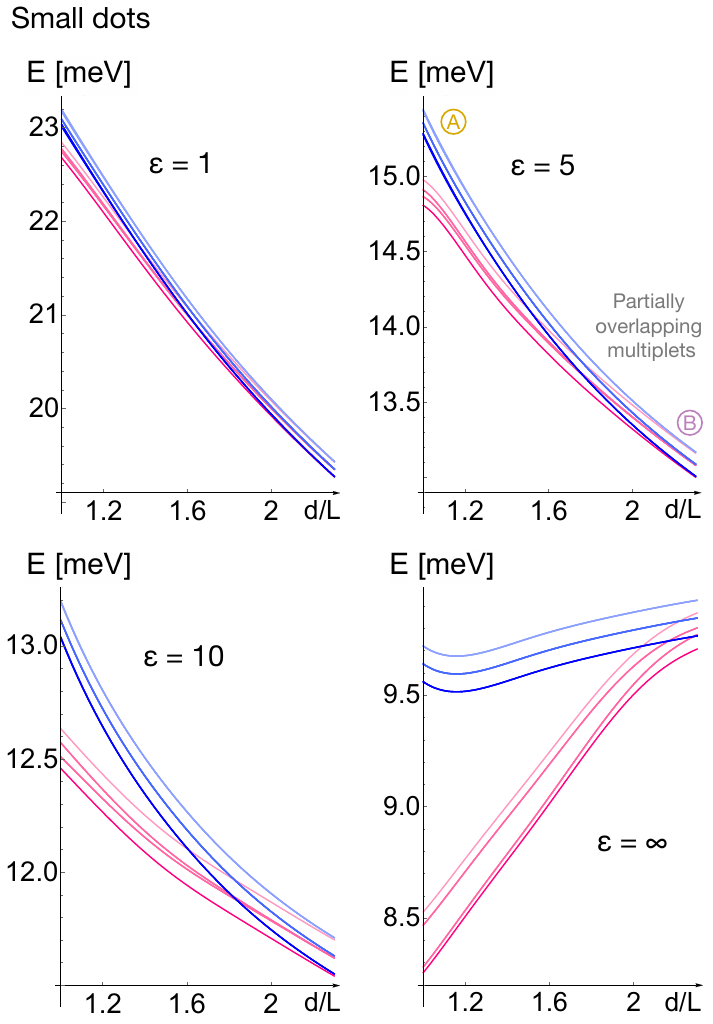}
    \caption{{Same as \fig{}\ref{fig:3} in the main text for different values of the dielectric constant, $\epsilon$. Two-particle spectra for $L=30$ nm and $\Delta_{0}=60$ meV.}
    }
    \label{fig:AppSpectraSmall}
\end{figure}

\begin{figure}[hb]
    \centering
    \includegraphics[width=0.99\linewidth]{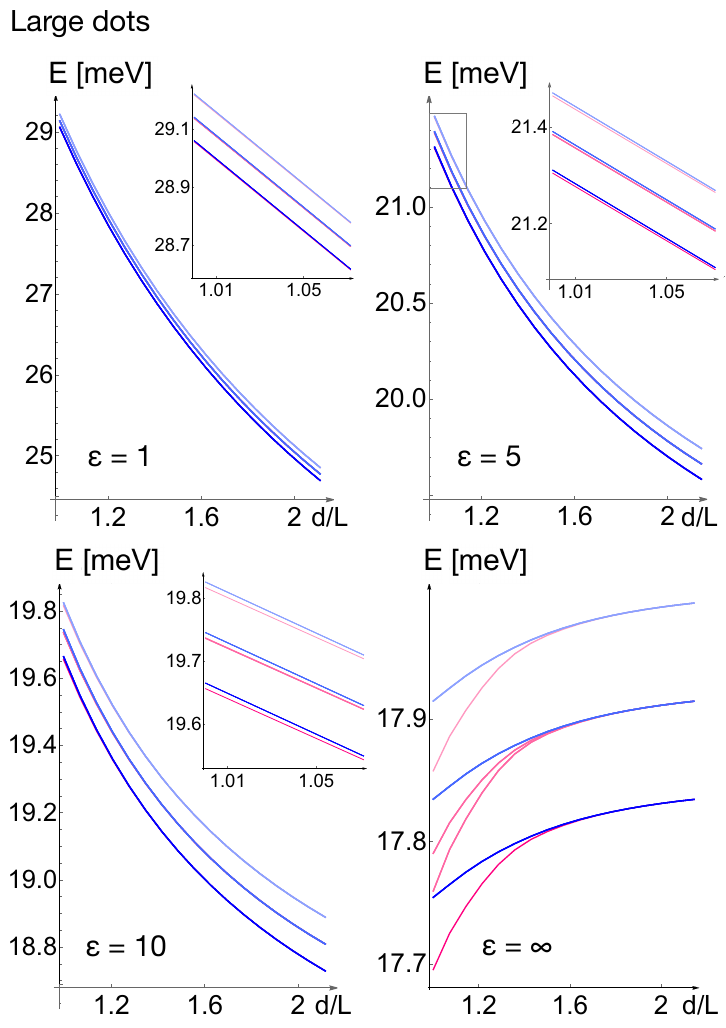}
    \caption{{Same as \fig{}\ref{fig:3} in the main text for different values of the dielectric constant, $\epsilon$. Two-particle spectra for $L=70$ nm and $\Delta_{0}=80$ meV.}
    }
    \label{fig:AppSpectraLarge}
\end{figure}

{Finally, \figs{}\ref{fig:AppSpectraSmall} and \ref{fig:AppSpectraLarge} illustrate the dependence of the low-energy \TP{} multiplets on the interaction strength tuned by the dielectric constant, $\epsilon$ (cf.~\fig{}\ref{fig:3} in the main text). Increasing the screening decreases the Hubbard parameters induced by the long-range Coulomb interactions, cf.~\fig \ref{fig:AppD607080}. The splitting between the orbitally symmetric and antisymmetric multiplet decreases with increasing interaction strength at fixed tunnelling (i.e., fixed inter-dot distance) as predicted by the dependence of the orbital splitting, $\Delta_{orb}$, \eqn{}\ref{eqn:Deltaorb}, on the Hubbard parameters. At low screening, there is an interplay between the long-range and short-range interactions and the tunneling. For sufficiently large screening, the splittings  are governed by $t$, and the short-range interactions.}
 
\clearpage
\section{Double Dot in a magnetic field}
\label{app:B}
In this appendix, we provide additional information concerning the \BLG{}\DQD{} in an external perpendicular magnetic field.

\begin{figure}[hb]
    \centering
    \includegraphics[width=0.99\linewidth]{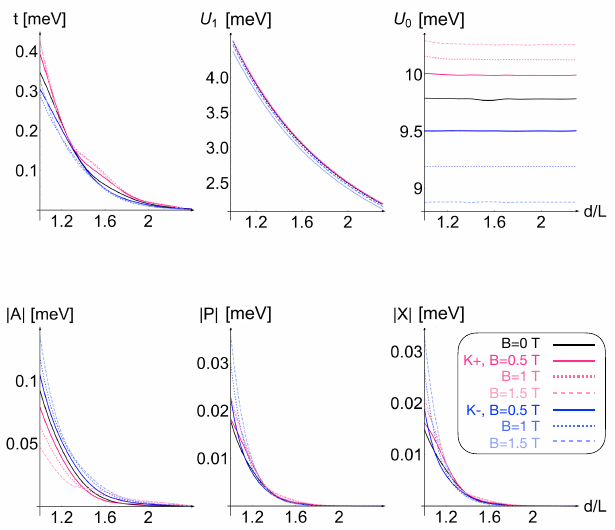}
    \caption{Hubbard parameters as a function of the inter-dot separation, $d$, for $\Delta_0=70$ meV and L$=50$ nm in the $K^{+}$ valley (magenta) and $K^{-}$  valley (blue) the for different B.
    }
    \label{fig:AppB}
\end{figure}

Figure \ref{fig:AppB} demonstrates the non-monotonic dependence of the Hubbard parameters $t$, $U_0$, $U_1$, $P$, $A$, and $X$ in the $K^{+}$ and $K^{-}$ valley on the magnetic field strength (cf.~\fig{}\ref{fig:5} in the main text).

\begin{figure}[ht]
    \centering
    \includegraphics[width=0.55\linewidth]{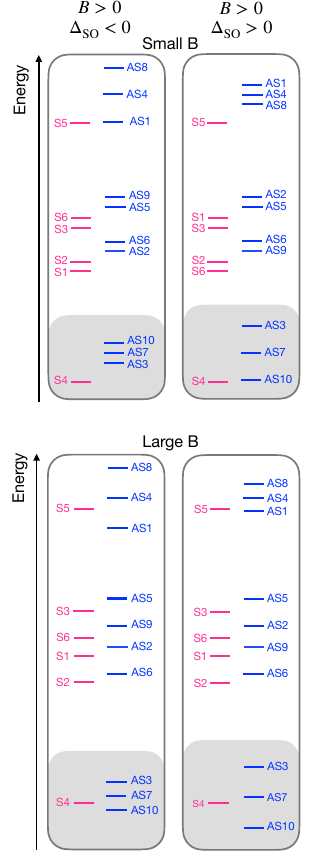}
    \caption{Level ordering of the \TP{} \DD{}  states in a finite magnetic field $B>0$ depending on the  sign  of the \SO{}-coupling gap and the external magnetic field strength. The gray shade highlights the $K^+$ valley-polarised low-energy states we focus on in the main text.
    }
    \label{fig:AppX}
\end{figure}

Figure  \ref{fig:AppX} illustrates the level ordering of all \TP{} \DD{}  states, \eqns{}\ref{eqn:StatesS} and \ref{eqn:StatesAS}, in a finite magnetic field $B>0$ depending on the sign of the \SO{}-coupling gap and the external magnetic field strength for the parameters chosen in the main text. 

\clearpage
\bibliography{QDLattice.bib}
\end{document}